\documentclass{aa}  

\usepackage{graphicx}
\usepackage{multicol,lipsum}
\usepackage{tabularx}
\usepackage{enumitem}
\usepackage{caption}
\usepackage{subcaption}
\usepackage{adjustbox}
\usepackage{makecell}
\usepackage{booktabs}
\usepackage{txfonts}
\usepackage{url}
\usepackage[colorlinks,citecolor=blue,urlcolor=blue,bookmarks=false,hypertexnames=true]{hyperref}
\usepackage [english]{babel}
\usepackage [autostyle, english = american]{csquotes}
\MakeOuterQuote{"}
\begin{document}

   \title{From traffic jams to roadblocks: The outer regions of TW Hya with ALMA Band 8}


   \author{S. Das\inst{1,2}
          \and
          N.T. Kurtovic\inst{3}
          \and
          M. Flock \inst{1}
          }

   \institute{Max Planck Institute for Astronomy, Königstuhl 17, 69117 Heidelberg, Germany\\
              \email{das@mpia.de}
         \and
             Department of Physics, Sardar Vallabhbhai National Institute of Technology, Dumas Road, 395007 Surat, India\\
        \and
            Max Planck Institute for Extraterrestrial Physics, Giebenbachstrabe 1, 85748, Garching, Germany
             }

   \date{Received ; accepted }

 
  \abstract
   {{Current ALMA surveys often underestimate protoplanetary disk sizes as many disks have extended low surface-brightness regions that fall below ALMA's detection limits. To effectively capture faint millimeter continuum emission in these outer regions, increased sensitivity is required.}}
   {In order to gain insights on the {connection between disk structure and} planet formation, we aim to uncover {continuum} emission in the outer regions of the disk around TW Hya. {Additionally, we aim to investigate the evolution of this disk by studying its dust properties.}}
   {We present Atacama Large Millimeter Array (ALMA) observations of TW Hya at 0.65 mm with $\sim$ 0.5 arcsecond angular resolution, together with high angular resolution archival {observations} at 0.87 mm, 1.3 mm, 2.1 mm and 3.1 mm. We constrain the outer disk emission {with both image-plane retrieval, and visibility-plane modeling with non-parametric and parametric fitting tools.}}
   {Our results confirm emission in the outer disk regions of TW Hya (60 au $\leq$ R $\leq$ 110 au) at 0.65 mm, 0.87 mm and 1.3 mm. {With image-plane retrieval}, we resolve the new continuum gap and ring, namely D79 and B86, at 0.87 mm and 1.3 mm. {With visibility-plane modeling}, we also detect this substructure at 0.65 mm in the form of a quasi-constant emission {at the 1$\sigma$ level}. {Furthermore, it} has a high spectral index of 3.7, which may indicate dust grain sizes $\ll$ 1 mm. It may be a dust trap or a traffic jam, that has a flux density of $\sim$ 60 mJy and a mass (1.59 $M_\oplus$) that accounts for up to 2\% of the dust disk at 0.65 mm.}
   {We confirm the existence of a faint ring in the outer regions of TW Hya at multiple millimeter wavelengths. {With visibility-plane modeling, we are able to set constrains that are 3 times better than the resolution of our Band 8 observations.}}


   \maketitle
%


\section{Introduction}

Planets are formed from the dust and gas in protoplanetary disks surrounding young stars. The study of these objects is key to understanding the physical processes involved in planet formation. More specifically, the study of the dust distribution in a disk allows us to test for mechanisms that drive planet formation and disk evolution \citep{manara2023, lesur2023}. Over the last decade, the Atacama Large Millimeter/sub-Millimeter Array (ALMA), one of the most powerful interferometers at (sub-)mm wavelengths, has allowed us to image disks at $\sim$ 5 au spatial resolutions and superb continuum sensitivity. These observational capabilities have revealed a large diversity of "substructures" in the dust continuum emission of protoplanetary disks, with the most common annular substructures being gaps and rings \citep{long2018, andrews2018}. Explanations for their origin range from planet-disk interactions, ice lines to {vertical shear instability (VSI)} and magneto-hydrodynamic zonal flows \citep{bae2023}. PDS 70 is one such example where we have found continuum rings as well as two directly imaged protoplanets \citep{keppler2018, haffert2019}. Nevertheless, a single planet can open multiple gaps, and not all gaps have a planet in them \citep{bae2017}. Some of the observed continuum rings in disks around HL Tau \citep{zhang2015} and HD 163296 \citep{zhang2021} can also be correlated to where one would expect ice-lines to be present in the midplane layer. \\

Another important observational metric in disks is their radial size. Due to radial drift, millimeter and centimeter sized dust grains are expected to migrate inward towards the high-pressure inner regions of disks, resulting in compact disks (\citealp{whipple}). However, observations reveal several extended dust disks because millimeter-sized grains persist in the outer disk regions for several million years (\citealp{testi,ricci2010}). Therefore, substructures must be present in these outer disks to prevent the large dust grains from getting drained rapidly. The presence of substructures, in turn, can be explained by the localized accumulation of pebbles within the gas pressure peaks of the disks \citep{pinilla2012}, subsequently triggering what is known as the streaming instability \citep{youdin}. We expect another phenomenon called viscous spreading to be responsible for extended dust disks \citep{rosotti2019}. When the bulk of the larger dust grains move inward to accrete onto the star, parts of the disk must move outward to conserve angular momentum. In principle, younger star forming regions should have more compact disks than older ones (e.g., \citealp{tazzari2017}). \\

Despite ALMA's excellent angular resolution and sensitivity, identifying and characterizing substructures is difficult. A limited angular resolution may leave substructures unresolved and a restricted sensitivity may result in the underestimation of disk sizes, especially in faint extended dust disks. The former has been the case for LkCa15 where initially only one ring was resolved at 0.25'' \citep{andrews2011b}, but later on \citealp{facchini2020} found the single ring to consist of two or three components with ALMA observations at a resolution of 0.05''. A similar consequence of low angular resolution observations is that multiple rings in a disk have the tendency to appear as a single rim \citep{parker}. In another study, \citealp{rosotti2019} predicted that current ALMA surveys do not possess enough sensitivity to accurately trace out the sharp outer edge of the millimeter dust continuum. Thus, precise characterization of the morphology of protoplanetary disk substructures is essential for gaining insight into the mechanisms that drive the evolution of these disks and may eventually lead to planet formation (e.g., \citealp{flock2015}, \citealp{bae2023}).\\

For our study, we choose TW Hya since it is the closest known protoplanetary disk, located only at a distance of 59.5 pc from Earth \citep{gaiacollab2018}. This makes it an optimal environment to study observational metrics that are influenced by the aforementioned physical mechanisms. Furthermore, TW Hya is characterized by 5 rings within the inner 50 au, which were identified through high-angular-resolution observations conducted in ALMA Band 7 \citep{huang2018a}, Band 6 \citep{tsukagoshi2019}, Band 4 \citep{tsukagoshi2016} and Band 3 \citep{macias2021} at the wavelengths of 0.87 mm, 1.3 mm, 2.1 mm and 3.1 mm respectively. It is a T Tauri star in the constellation of Hydra. TW Hya has a spectral classification of M0.5 with an effective temperature $T_{eff}$ = 3800 K, a mass of 0.6 M$_{\odot}$, and an age of 8 Myr \citep{sokal2018}. The star is still actively accreting from its disk at a rate of $\dot{M}$ = 4-20 $\times$ 10$^{-10}$  M$_{\odot}$yr$^{-1}$ \citep{herczeg2004}. It has an inclination angle of {5\textdegree} \citep{huang2018a} and thus, it is nearly face-on. \\

The spectral energy distribution (SED) curve reveals that the TW Hya disk is a "transition disk" \citep{strom1990} and the system consists of two parts - an optically thin, high-temperature and dust free inner region spanning from 0.06 to 1.3 au and a cold outer disk stretching out to 200 au \citep{hughes2007}. A study conducted in scattered light with HST/STIS indicates emission in the TW Hya disk up to 180 au and the presence of a wide gap at 85 au (\citealp{debes2017}). Similarly, another study conducted with VLT/SPHERE reports a gap at {90.5 au} (\citealp{boekel2017}). However, high angular resolution ALMA observations of the disk have reported emission extending only up to $\sim$ 60 au both in the continuum images and their intensity profiles \citep{huang2018a, macias2021}. \citealp{rosotti2019} {suspects} that a low sub-mm opacity is responsible for not being able to estimate the accurate disk sizes with limited ALMA sensitivity. Recently, \citealp{ilee} proposed the presence of an outer gap and ring centered at 82 au and 91 au respectively, using observations from ALMA Band 7 ($\sim$ 294 GHz). Therefore, it becomes reasonable to believe that the "outer disk" (R $\geq$ 60 au) could be visible at other ALMA bands. Particularly, a short wavelength could effectively probe its small dust grains. \\

In this work, we present $\sim$ 0.5 arcsecond (30 au) observations at ALMA Band 8 (0.65 mm; $\sim$ 460 GHz). Despite the low angular resolution of the data, it has a high S/N and a high sensitivity, allowing us to provide insights into the characteristics of TW Hya's outer disk. We analyze this data together with high angular resolution archival data from Band 3, Band 4, Band 6 and Band 7 to set constraints on the outer disk properties. In Section \ref{2}, we provide details about the datasets and their calibration procedures. In Section \ref{3}, we discuss the various {techniques} implemented to analyze our observations and their results. In Section \ref{4}, we discuss the possible explanations for our results and suggest further studies. The paper concludes with a short summary in Section \ref{5}. 



\section{Observations and data reduction} \label{2} 
Our work analyzes data from Bands 3 (3.1 mm), 4 (2.1 mm), 6 (1.3 mm), 7 (0.87 mm) and 8 (0.65 mm). The observations in the first four bands have previously been processed and analyzed by \citealp{macias2021}, \citealp{tsukagoshi2016}, \citealp{tsukagoshi2019}, and \citealp{huang2018a}. TW Hya was observed in ALMA Band 8 (0.65 mm) during three distinct execution phases comprising Cycle 4 (27th March, 2017) and Cycle 5 (23rd May and 5th June, 2018) with the configurations C40-1 and C43-2 respectively (Project code: 2016.1.01399.S; PI: V. Salinas). The data was initially calibrated using the Common Astronomy Software Applications package v5.1.1 (\texttt{CASA}; \citealp{McMullin2007}) with the script and pipeline provided by the ALMA team. For each observation, the baselines ranged from 15 - 161 m, 15 - 314 m and 15 - 331 m with 42, 46 and 47 antennas respectively. The total observing time was 146 minutes with an on-source time of 48.7 minutes. J1229+0203 was used for flux and bandpass calibration whereas J1037-2934 was used for flux and phase calibration. The central frequencies for the continuum were 456 GHz, 458.7 GHz, 469.4 GHz, 470.3 GHz and 470.4 GHz. \\

To improve the quality of the ALMA Band 8 dataset, we applied self-calibration following a strategy similar to \citealp{andrews2018}. Self calibration and further processing of the data was done using \texttt{CASA} v6.4.1. To extract the continuum, channels within $\pm$ 25 km s$^{-1}$ around the emission lines were flagged. The targeted line was NH$_2$D at 456 GHz, 469.4 GHz, 470.3 GHz and 470.4 GHz.  An averaging over a width of 125 MHz was applied to all channels. We then changed the position of the phase center to the center of the observation, and redefined the coordinate system to match the observations in Band 6 and Band 7 from \citealp{huang2018a}. Therefore, the phase centers of the visibilities of each of the three datasets were corrected to J2000 11h01m51.817995s -34d42m17.234924s with the \texttt{fixplanets} task. There is a flux variation among the ALMA Band 8 observations with a scatter of about 5\%. To combine these datasets, we rescaled the fluxes of the second and third dataset to match the flux of the first one, which has a flux close to the average of the three. This correction is smaller than the 10\% uncertainty in the total flux calibration considered in our analysis of the spectral index (see Section \ref{3.4.2}). Finally, the three datasets were concatenated. The self-calibration consisted of 4 rounds of phase calibrations and 1 round of amplitude calibration. For the phase calibration, solution intervals of infinite, 180 s, 90 s and 60 s were used. The amplitude calibration had a solution interval of 240 s. In each step, the images were cleaned down to 1$\sigma$. The final dataset\footnote{The data associated with this paper is publicly available and can be found in \url{https://doi.org/10.5281/zenodo.12677551}} showed a S/N improvement of $\sim$ 36\% compared to the initial dataset. \\

\begin{figure*}
    \sidecaption
    \includegraphics[width=12.9cm]{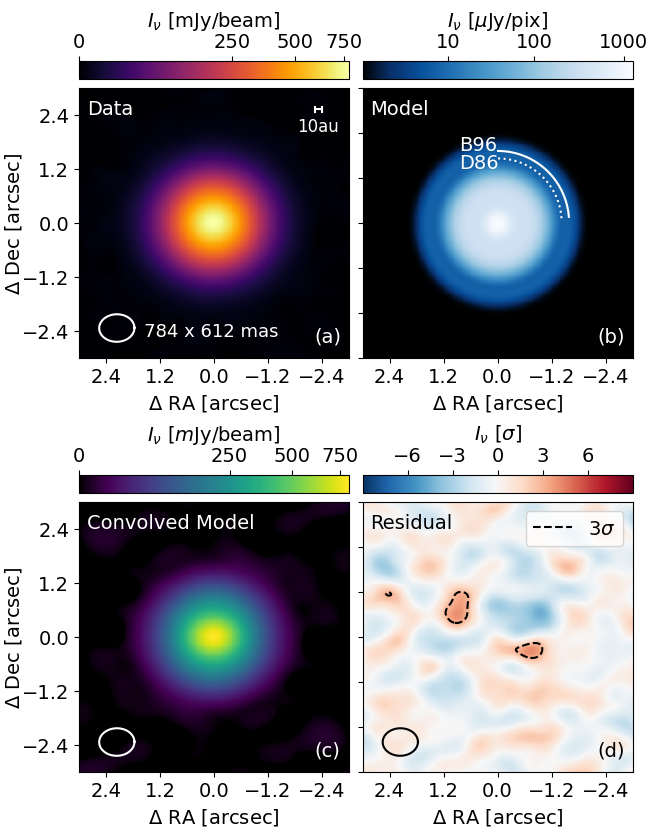} 
    \caption {{Results of parametric modeling: (a) Image of the dust continuum emission of TW Hya's protoplanetary disk at a wavelength of 0.65 mm. The ellipse at the bottom left corner shows the beam (dimensions : 0.784'' $\times$ 0.612''; PA = 91.04\textdegree) used to convolve the image. The bar at the upper right corner shows the spatial scale which is equivalent to 10 au. (b) Unconvolved best-fit model of the 0.65 mm dust continuum emission fit with 3 gaussians using MCMC and \texttt{galario} (see Section \ref{3.2.2}). The substructures (D86 and B96) obtained from the intensity profile (see Figure \ref{fig:mcmc_profile}) are shown with dotted and solid curves respectively. (c) Best-fit model that has been convolved with the same beam as the image. (d) The residuals of the emission as multiples of the rms ($\sigma$). Residuals at the level of 3$\sigma$ are shown with contours. It was obtained when the model was directly subtracted from the data. The beam used to generate this has the same dimensions as that used for the image.}}
    \label{fig:mcmc_residual}
\end{figure*}

\section{Results}\label{3} 

\subsection{Continuum image of TW Hya at 0.65 mm}

The final continuum image was made with the \texttt{tclean} task in CASA v6.4.1. After testing several Briggs weighting values, it was determined that a \textit{robust} of 0.5 gave the optimal balance between sensitivity and angular resolution. The total image size and pixel size were set to 10.24'' and 0.02'' (i.e. $\sim$ 0.04 times the angular resolution of the observation) respectively. We cleaned using multi-scale, with components of 0'', 0.06'', 0.16'' and 0.24''. The synthesized beam has the dimensions 0.784'' $\times$ 0.612'' (47  $\times$ 37 au) and a position angle of 91.04\textdegree. This resulting image is shown in Figure \ref{fig:mcmc_residual} (a). The peak intensity of the source is 780.53 mJy beam$^{-1}$. Using a circular mask of 2.5'' (150 au)\footnote{The mask size was conservatively set to include all of the millimeter emission of TW Hya, based on our visibility modeling results described in Section \ref{3.2}}, the total integrated flux density is 3029.85 $\pm$ 0.79 mJy where the uncertainty was calculated as the rms multiplied by the square root of the number of beams in the mask. The rms is 0.22 mJy beam$^{-1}$ (0.14 mJy arcsec$^{-2}$), measured in an annulus of 2.6'' and 5.0'' where the image is signal-free. Lastly, the peak signal-to-noise ratio (S/N) of our observation is 3638.64. The 459 GHz continuum flux is 2 times brighter than the 347 GHz flux (1542 $\pm$ 3 mJy) and 5 times brighter than the 233 GHz flux (575.4 $\pm$ 0.9 mJy) measured with ALMA Band 7 and Band 6 respectively in \citealp{macias2021}. Due to the low angular resolution of the data, we do not resolve any substructures in the image plane. \\


\subsection{Characterizing the continuum emission at 0.65 mm} \label{3.2}

One way to characterize the emission in observations is by creating one-dimensional intensity profiles. On azimuthally averaging the intensity obtained from the continuum image, we were not able to detect any substructures. Hence, we utilized the visibility fitting techniques - the \texttt{Frankenstein} code (\citealp{jennings2020} v1.1.0), which fits a non-parametric one-dimensional model to the visibilities, and the \texttt{galario} code (\citealp{tazzari2018}), which we used with \texttt{emcee} (\citealp{emcee}) to fit a parametric model to the visibilities. \\

In the following sections, we characterize the substructures by their location, width and contrast ratio. The gap and ring locations are defined as the radial positions where the intensity profile reaches a local minima followed by a local maxima respectively. The prefix "B" and "D" are used to represent a "Bright" Ring and "Dark" Gap respectively. The prefix is followed by the substructure's radial location rounded to the nearest whole number \citep{dsharp2}. The width is defined as the difference in radial locations where the intensity is equal to the average intensity between the ring maxima and the gap minima. For each gap-ring pair, we will obtain three such radial locations. Accordingly, the gap width is calculated as the difference between the first and the second radial locations, and the ring width is calculated as the difference between the second and the third radial locations. The contrast ratio is defined as the intensity ratio between the ring maxima and the gap minima. The uncertainty is calculated from the radial location where we obtain the same intensity as the ring maxima or the gap minima in the upper and lower limits of the uncertainty in the emission. To calculate the disk sizes, we constructed cumulative intensity profiles \citep{tripathi} from our azimuthally averaged intensity profiles. We also calculated the uncertainty in disk sizes based on the uncertainty in our intensity profiles. Disk sizes are usually estimated as the radius enclosing 68\%, 95\% or 99\% of the total flux density.

\subsubsection{{Non-parametric modeling} with frank} \label{3.2.1}

The \texttt{Frankenstein} code (abbreviated as \texttt{frank}) fits the visibilities to reconstruct the brightness profile of a protoplanetary disk. The model is azimuthally symmetric by construction. Furthermore, it assumes that the emission originates from a geometrically thin and optically thick disk thereby, accommodating the source's inclination and position angle. We generated several profiles with the \texttt{FrankFitter} task using both the "Normal" and "logNormal" methods. We used the following hyper-parameters: $\alpha$ = \{1.05, 1.2, 1.3\},  w$_{smooth}$ = \{10$^{-1}$, 10$^{-4}$\}, $R_{max} = 5.0''$, $p_0 = 10^{-35}$ with 300 radial points. We subtracted 20 mJy from the real part of our visibilities, assuming it to be the flux density of {the} point source. The brightness profiles thus obtained are shown in Appendix \ref{appendix_A}. We find that {the logNormal fit with} $\alpha$ = 1.3 and w$_{smooth}$ = 10$^{-4}$ gives the most {consistent} results\footnote{{We find that the intensity profile created with our parametric model (Figure \ref{fig:mcmc_profile}) closely resembles the profile generated with these parameters in frank's logNormal mode.}}. {\citealp{jennings2020} discusses that the $\alpha$ parameter controls the S/N threshold and therefore, a higher value fits less of the noisiest data. Additionally, a smaller w$_{smooth}$ avoids smoothing the power spectrum too strongly\footnote{\url{https://discsim.github.io/frank/tutorials/prior\_sensitivity.html}}}.  \\ 

 We direct our focus {to} the logNormal profiles since they contain fewer noise-induced artifacts. In all our logNormal profiles, we obtain a shoulder-like emission between 80 au and 97 au. The brightness profile reconstructed beyond 125 au corresponds to a flux density that is 0.003 times the total flux density of the disk and therefore, the rings do not signify real emission. The most important observation is that the flux density (63.15 $\pm$ 13.99 mJy) obtained between 80 au and 97 au is robust against changes in the $\alpha$ and $w_{smooth}$ values. Particularly, we find that the w$_{smooth}$ parameter influences the morphology of the shoulder, with higher values smoothing down the quasi-constant emission. The radial position of the shoulder-like emission is consistent with the outer disk substructures proposed by \citealp{ilee}, who also used \texttt{frank} to recover this emission. In their work, they identified that dust continuum emission in TW Hya's protoplanetary disk at $\sim$ 294 GHz extends up to $\sim$ 100 au, and the disk has a gap and a ring centered at 82 au and 91 au respectively. \\

\subsubsection{{Parametric modeling} with MCMC and galario} \label{3.2.2} 

\begin{figure*}
    \centering
    \includegraphics[width=0.48\textwidth]{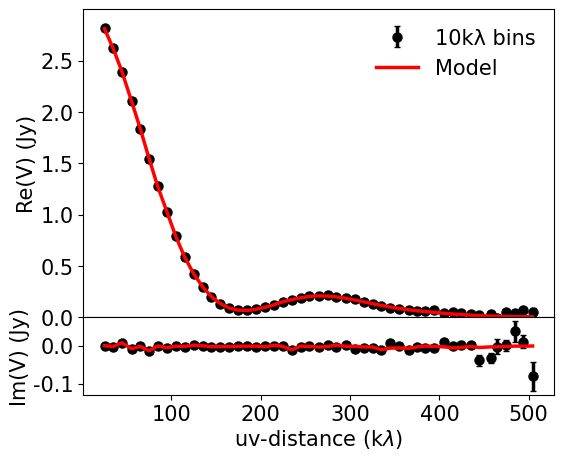} 
    \includegraphics[width=0.515\textwidth]{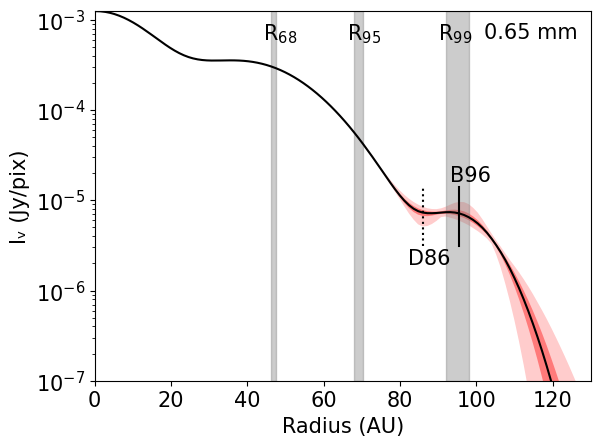}
    \caption{{Results of parametric modeling}: \textit{(Left)} The re-centered and deprojected visibilities of the 0.65 mm observation binned to 10 k$\lambda$, where the top row shows the real part and the bottom row shows the imaginary part. The visibilities of the best fit model is shown in red. \textit{(Right)} Intensity profile of the TW Hya protoplanetary disk at 0.65 mm made with 5000 random posteriors of our MCMC sample. The median intensity profile is shown with a black curve, whereas the 1$\sigma$ and 3$\sigma$ uncertainties are shown with dark and light red shaded regions respectively. The local minima and maxima representing a dark gap and a bright ring at 86 au and 96 au are shown with dotted and solid lines respectively. The disk size estimates are shown as $R_{68}$, $R_{95}$ and $R_{99}$ in the form of gray columns.}
    \label{fig:mcmc_profile}
\end{figure*}

To confirm the existence of the outer disk substructures suggested by \texttt{frank}, we also modeled the disk's visibilities with the Markov Chain Monte Carlo (MCMC) algorithm (\texttt{emcee} package) and the \texttt{galario} package. The galario package is used to generate synthetic visibilities for a model image. In our model, we assumed that the TW Hya disk is azimuthally symmetric. Based on the frank-fit, we found that the 0.65 mm disk can be defined with 1 centrally peaked Gaussian and 2 Gaussian rings. The functional form of a Gaussian is given by,  
\begin{equation} \label{Equation1}
I(r) = {I_{i}\:exp\:(-\frac{(r-R_{i})^2}{2\sigma_{i}^2}}), \\
\end{equation}
\noindent
where, i = \{0, 1, 2\} denotes the index of the Gaussian component, $r$ is the radial distance from the source, $I_{i}$ is the peak intensity, $R_{i}$ is the position of the centre of the peak intensity (or, ring location) and $\sigma_{i}$ is the standard deviation. Since the first Gaussian is centrally peaked, we set $R_{0}$ = 0. Therefore, we fit the parameters I$_0$, $\sigma_0$, I$_1$, $R_1$, $\sigma_1$, I$_2$, $R_2$, and $\sigma_2$. We also fit the right ascension offset ({$\Delta$R}) and the declination offset ({$\Delta$D}) between the observation phase centre and the centre of the disk. We considered the inclination angle (5.8\textdegree) and the position angle (151.6\textdegree) from the high angular resolution studies \citep{huang2018a}, and did not fit them. Thus, a total of 10 parameters were fit with the MCMC algorithm. The results of our fitting along with the staircase plot of our samples are shown in Appendix \ref{appendix_B}. From our samples, we defined the best-fit model as the one having the maximum likelihood. The data along with the best-fit model and its residuals are shown in Figure \ref{fig:mcmc_residual}. We also show the unconvolved and convolved models separately. Furthermore, we generated azimuthally averaged intensity profiles by calculating the median, 1$\sigma$ and 3$\sigma$ intensities for 5000 random posteriors of our sample. The intensity profile along with the deprojected and re-centered visibilities from the best-fit model are {shown} in Figure \ref{fig:mcmc_profile}. \\

The ratio of emission amplitudes of the residual and the data is $\sim$ 0.0013. The model describes 99.88\% of the original peak intensity showing that our axisymmetric model is well suited to describe the 0.65 mm disk. The outer ring in the median intensity profile appears in the form of a quasi-constant emission and peaks at $\sim$ 96 au. This ring has a standard deviation ($\sigma_{2}$) of $\sim$ 8.38 au. Although we do not resolve the separation between the main disk emission and the outer ring to the 1$\sigma$ uncertainty, we observe a local minima at $\sim$ 86 au when we consider the shape of the 1$\sigma$ and 3$\sigma$ uncertainties. As a result of the low intensity, the gap and ring locations have large uncertainties (as large as $\sim$ 0.2 times the angular resolution). We calculated the ring width as the Full Width {at} Half Maximum (FWHM)\footnote{$FWHM = 2\sigma\sqrt{\:2\: ln(2)} = \:\sim2.355\:\sigma.$} of the Gaussian describing the outer ring. This gives us a ring width of 19.73 $\pm$ 2.44 au. \\

The outer ring has a flux density of $59.24^{+14.64}_{-10.31}$ mJy, and the model disk has a total flux density of $3075.11^{+71.55}_{-68.44}$ mJy. The uncertainties are calculated at the level of 1$\sigma$. {The integrated flux density is consistent with the Spectral Energy Distribution (SED) for TW Hya compiled by \citealp{menu2014} when implementing models from \citealp{calvet2002}, \citealp{ratzka2007} and \citealp{andrews2012}. The B96 flux density also closely aligns with the flux density of the shoulder obtained with the logNormal frank brightness profiles.} {Additionally,} we calculated the following disk sizes: $R_{68}$ = 46.85 $\pm$ 0.61 au, $R_{95}$ = 69.13 $\pm$ 1.20 au and $R_{99}$ = 95.0 $\pm$ 3.0 au. Even though our model gives a tentative gap and ring location, higher angular resolution ALMA Band 8 observations (at least $\sim$ 0.1'', 6 au), with {high} sensitivity and S/N, are required to resolve the separation and morphology of this substructure.

\subsection{{Image-plane retrieval with} intensity profiles of the continuum emission at 3.1mm, 2.1mm, 1.3mm and 0.87mm} \label{3.3}

\begin{table*}
\caption{Imaging results at Bands 3, 4, 6, 7 and 8}
\label{table:1}
\centering
\begin{tabular}{c c c c c c c}     
\hline\hline       
\thead{Band} &  \thead{Central \\ Freq. (GHz)} &  \thead{Wavelength (mm)} &  \thead{Beam \\ Dimensions} &  \thead{rms \\ ($\mu$Jy beam$^{-1}$)} &  \thead{Peak S/N} &  \thead{Total Flux \\ Density (mJy)\tablefootmark{a}} \\ [3ex]

\midrule

\vspace{2mm}
   3 & 97.5 & 3.08 & 0.05'' $\times$ 0.05'', PA = 0.0\textdegree & 5.3 & 137.26 & 52.97 $\pm$ 0.21 \\  
\vspace{2mm}
   4 & 145.0 & 2.07 & 0.05'' $\times$ 0.05'', PA = 0.0\textdegree & 11.0 & 85.21 & 150.89 $\pm$ 0.44 \\ 
\vspace{2mm}   
   6 & 233.0 & 1.29 & 0.05'' $\times$ 0.05'', PA = 0.0\textdegree & 7.1 & 349.23 & 590.32 $\pm$ 0.28 \\ 
\vspace{2mm}   
   7 & 346.8 & 0.865 & 0.05'' $\times$ 0.05'', PA = 0.0\textdegree & 23.9 & 213.18 & 1559.72 $\pm$ 0.96 \\ 
   8 & 458.6 & 0.654 & 0.784'' $\times$ 0.612'', PA = 91.04\textdegree & 220.0 & 3838.64 & 3029.85 $\pm$ 0.79 \\
\midrule     
\vspace{3mm}
\end{tabular}
\newline Notes: \tablefoottext{a}{The uncertainty is calculated as the rms multiplied by the square root of the number of beams in the mask. $\sim$ 10\% flux calibration uncertainty is not included.}
\vspace{2mm}
\end{table*}

\begin{figure*}
    \centering
        \includegraphics[width=0.49\textwidth]{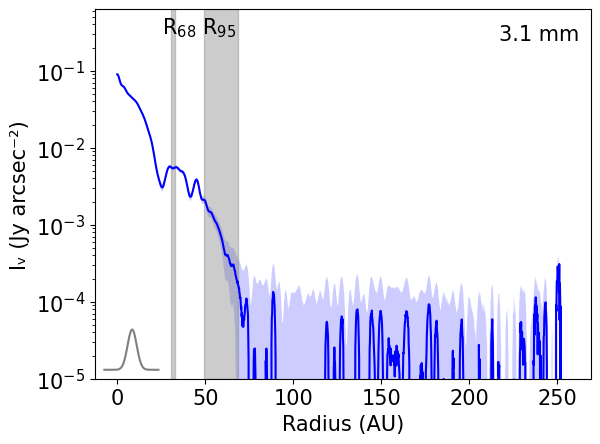}      
        \includegraphics[width=0.49\textwidth]{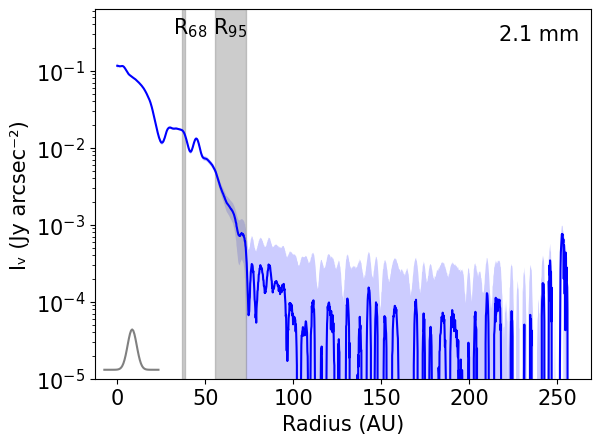} 
        \includegraphics[width=0.49\textwidth]{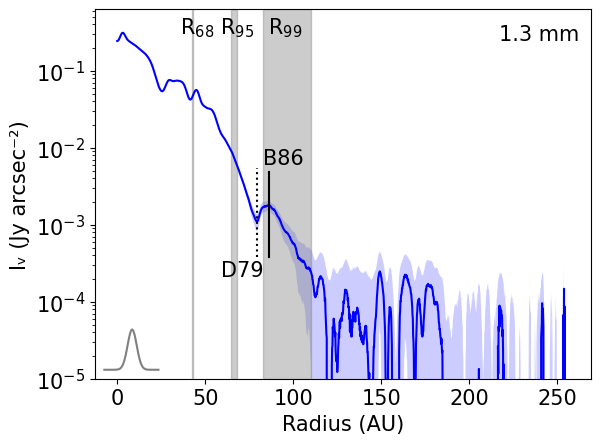} 
        \includegraphics[width=0.49\textwidth]{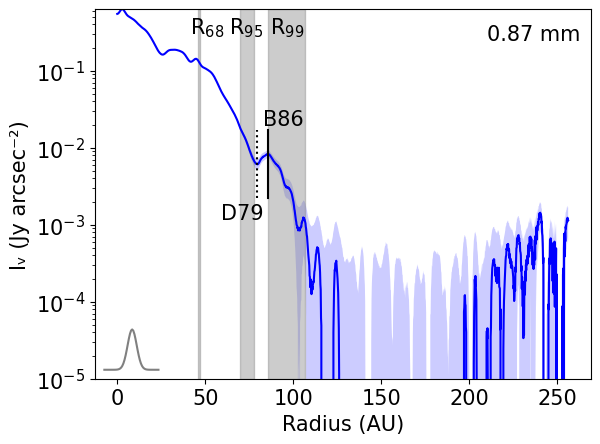} 
        \vspace{4mm}
    \caption{{Results of image-plane retrieval: Intensity profiles of the TW Hya protoplanetary disk at 3.1 mm (top-left), 2.1 mm (top-right), 1.3 mm (bottom-left) and 0.87 mm (bottom-right). The images from which they were generated were smoothed to have the resolution of circular beams (0.05'' $\times$ 0.05''; PA = 0.0\textdegree) (see Table \ref{table:1}). The bottom left corner of each panel shows the profile of the synthesized beam. The shaded blue region represents the uncertainty in the emission, and were calculated with equations \ref{Equation2} and \ref{Equation3}. The 1.3 mm and 0.87 mm intensity profiles contain the outer-disk substructures D79 and B86 which are shown with black dotted and solid lines respectively. The 3.1 mm and 2.1 mm intensity profiles do not reveal any outer-region emission to the noise level. The $R_{68}$, $R_{95}$ and $R_{99}$ disk size estimates are shown with gray columns. We do not show the $R_{99}$ disk size for the 2.1 mm and 3.1 mm profiles because it coincides with $R_{95}$ due to lower {surface} brightness levels.}}
    \label{fig:macias_profiles}
\end{figure*}

\begin{table*}
\caption{Properties of outer-disk substructures at 1.3 mm, 0.87 mm and 0.65 mm}             
\label{table:outer_disk}      
\centering   
\begin{tabular}{c c c c c c c}     
\hline\hline       
\thead{Band} &  \thead{Wavelength (mm)} & \thead{Gap \\ Location (au)}  &  \thead{Gap \\ Width (au)} &  \thead{Ring \\ Location (au)} &  \thead{Ring \\ Width (au)} &  \thead{Contrast \\ Ratio} \\ [3ex]

\midrule

\vspace{2mm}
   6 & 1.29 & $79.11^{+1.80}_{-1.80}$ & $4.51 ^{+1.26}_{-1.26}$ & $86.33^{+4.87}_{-4.87}$ & $9.02^{+1.26}_{-1.26}$ & $1.61^{+0.96}_{-0.96}$  \\  
\vspace{2mm}
   7 & 0.865 & $79.47^{+1.98}_{-1.98}$ & $4.69^{+1.44}_{-1.44}$ & $85.61^{+1.98}_{-1.98}$ & $6.31^{+1.44}_{-1.44}$ &  $1.37^{+0.96}_{-0.96}$ \\ 
   8\tablefootmark{a} & 0.654 & $85.97^{+6.02}_{-2.40}$ & -\tablefootmark{b} & $95.59^{+2.40}_{-6.01}$ & $19.73^{+2.44}_{-2.44}$ & $1.05^{+0.62}_{-0.62}$ \\ 
   
\midrule     
\end{tabular}
\newline 
\flushleft
Notes: \tablefoottext{a}{The 0.65 mm substructure properties are calculated from the intensity profile in Figure \ref{fig:mcmc_profile}. \\}
\hspace{8mm} \tablefoottext{b}{We do not report the D86 gap width since the separation is not clearly resolved in our model.}
\vspace{1mm}
\end{table*}

For comparison with our 0.65 mm (Band 8) intensity profile, we utilized the images generated by \citealp{macias2021} with ALMA observations at 3.1 mm (Band 3), 2.1 mm (Band 4), 1.3 mm (Band 6) and 0.87 mm (Band 7). The image properties are shown in Table \ref{table:1}. It is important to note that these images were smoothed to have the resolution of circular beams (0.050'' $\times$ 0.050''; PA = 0.0\textdegree). We used the method described in Appendix \ref{appendix_C} to generate azimuthally averaged intensity profiles at these 4 wavelengths. The profiles obtained are shown in Figure \ref{fig:macias_profiles}. \\

Within 0 - 60 au, we are able to resolve the substructures at all four wavelengths that were previously obtained by \citealp{dsharp2} (their Table 1) and \citealp{macias2021}. Beyond 60 au, we also find the substructures D79 and B86 at 0.87 mm and 1.3 mm\footnote{{For reference, the gap and ring are not detected at 1$\sigma$ in the image plane at 0.87 mm and 1.3 mm, and can only be resolved with the azimuthally averaged intensity profiles.}}. At both the wavelengths, we find that the lower limit of the uncertainty in the emission reaches zero at $\sim$ 110 au. Beyond this limit, noise starts dominating the profiles. The characteristics of the outer disk substructures at 1.3 mm, 0.87 mm and 0.65 mm are detailed in Table \ref{table:outer_disk}. We also estimated the $R_{68}$, $R_{95}$ and $R_{99}$ disk sizes, which are shown in Table \ref{table:size_flux}. The circular beams contribute to lowering the noise level, which otherwise dominates the outer disk region when the images are resolved with an elliptical beam. This is explained by the circular beam being larger, and thus including more emission in a single resolution element. At 0.87 mm and 1.3 mm, these circular beam images are found to have a $\sim$ 150\% increase in peak S/N and a $\sim$ 50\% decrease in rms than the images convolved with an elliptical beam (see \citealp{macias2021} Table 1). The presence of the outer disk at 1.3 mm, 0.87 mm and 0.65 mm is especially interesting since it may indicate the presence small dust grains in the outer disk regions. However, we do not recover any outer-disk emission at 2.1 mm and 3.1 mm to the noise level. Signal dominated emission at these wavelengths extends only upto $\sim$ 70 au, similar to the disk sizes reported by \citealp{andrews2016_twhya}, \citealp{huang2018a} and \citealp{macias2021}. If there are larger sized mm dust grains present in the outer disk, our data lacks the S/N and sensitivity required to capture it. We find that the Band 3 and Band 4 images have a 60\% - 70\% lower peak S/N compared to the Band 6 and Band 7 images at the same angular resolution. However, if the large grains are indeed absent, radial drift could be responsible for causing an inward drift of millimeter sized dust grains from the outer disk regions \citep{whipple}, especially when the star is still actively accreting \citep{herczeg2023}. \\      

Considering that the angular resolution of the 0.65 mm observation is 0.5'' (30 au), {shifts in the locations of the substructures at various wavelengths are $\sim$ 0.3 times the angular resolution, and therefore, they are within the acceptable range}. Accounting for the uncertainty in the ring width, the width of the B86 ring at 0.87 mm and 1.3 mm are consistent. Our current estimates {also} tell us that the outer ring {at 0.65 mm} may be wider than the outer ring at 0.87 mm and 1.3 mm.    

\subsection{Properties of the outer-disk substructures} 


A multi-wavelength study allows us to highlight the characteristics of TW Hya's outer-disk at various wavelengths. Some of those properties have been discussed in the sections below.

\subsubsection{Spectral index} \label{3.4.2}

\begin{table*}
\caption{Disk sizes, outer ring flux and mass at 3.1 mm, 2.1 mm, 1.3 mm, 0.87 mm and 0.65 mm}         
\label{table:size_flux}      
\centering   
\begin{tabular}{c c c c c c c c}     
\hline\hline       
\thead{Band} &  \thead{Wavelength (mm)} & \thead{Frequency (GHz)}  &  \thead{$R_{68}$ (au)\tablefootmark{a}} &  \thead{$R_{95}$ (au)\tablefootmark{b}} &  \thead{$R_{99}$ (au)\tablefootmark{c}} &  \thead{Flux of \\ Outer Ring (mJy)\tablefootmark{d}} & \thead{Mass of \\ Outer Ring ($M_\oplus$)\tablefootmark{e}} \\ [3ex]

\midrule
\vspace{2mm}
   3 & 3.08 & 97.5 & 31.76 $\pm$ 1.18 & 58.91 $\pm$ 9.56 & - & 0.06 $\pm$ 0.09 & 0.04 $\pm$ 0.06 \\
\vspace{2mm}
   4 & 2.07 & 145.0 & 37.62 $\pm$ 0.91 & 64.32 $\pm$ 9.03 & - & 1.53 $\pm$ 0.18 & 0.43 $\pm$ 0.05 \\
\vspace{2mm}
   6 & 1.29 & 233.0 & 42.85 $\pm$ 0.18 & 66.21 $\pm$ 1.71 & 96.52 $\pm$ 13.62 & 10.7 $\pm$ 0.12 & 1.08 $\pm$ 0.01 \\  
\vspace{2mm}
   7 & 0.865 & 346.8 & 46.46 $\pm$ 0.37 & 73.61 $\pm$ 4.06 & 96.34 $\pm$ 10.55 & 45.38 $\pm$ 0.44 & 2.04 $\pm$ 0.02 \\ 
   8 & 0.654 & 458.6 & 46.85 $\pm$ 0.61 & 69.13 $\pm$ 1.20 & 95.00 $\pm$ 3.00 & 59.24 $\pm$ 14.64 & 1.59 $\pm$ 0.39 \\ 
   
\midrule     
\vspace{3mm}
\end{tabular}
\flushleft
Notes: The disk sizes are calculated from cumulative radial profiles \citep{tripathi}, which in turn are calculated from the azimuthally averaged intensity profiles. The 3.1 mm, 2.1 mm, 1.3 mm and 0.87 mm disk sizes are calculated using the intensity profiles in Figure \ref{fig:macias_profiles}, whereas the {0.65 mm} disk sizes are calculated from the intensity profile in Figure \ref{fig:mcmc_profile}. We do not report $R_{99}$ for the 2.1 mm and 3.1 mm profiles because {it coincides} with $R_{95}$ due to lower {surface} brightness levels. 
\newline \tablefoottext{a}{The radius enclosing 68\% of the total flux density.\\}
\tablefoottext{b}{The radius enclosing 95\% of the total flux density. \\}
\tablefoottext{c}{The radius enclosing 99\% of the total flux density. \\}
\tablefoottext{d}{The 3.1 mm, 2.1 mm, 1.3 mm and 0.87 mm outer ring fluxes are calculated with the {\texttt{imstat} task in CASA with masks at} 79 au and 93 au, whereas the 0.65 mm one is calculated from the {best-fit} MCMC model. \\}
\tablefoottext{e}{The mass of the outer ring in multiples of the mass of the earth. Refer to Section \ref{3.4.3} for further details.}
\vspace{2mm}
\end{table*}

\begin{figure*}
    \centering
    \includegraphics[width=0.49\textwidth]{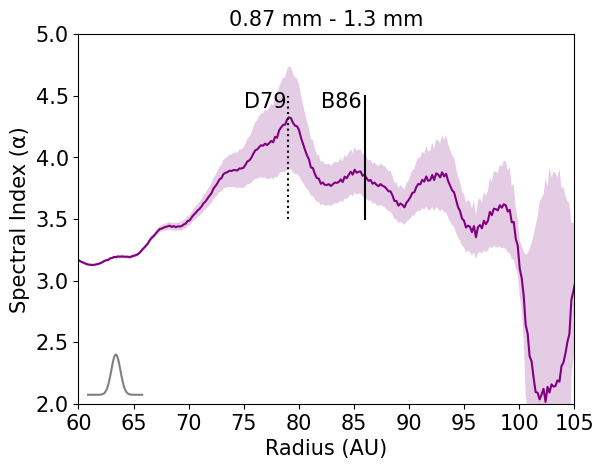}
    \includegraphics[width=0.49\textwidth]{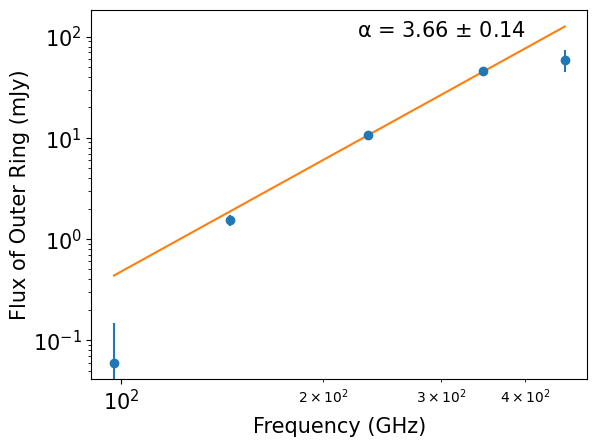}
    \vspace{1mm}
    \caption{{\textit{(Left)} 0.87 mm - 1.3 mm spectral index profile of TW Hya's outer disk (R $\geq$ 60 au). This is plotted with the intensity profiles in Figure \ref{fig:macias_profiles}. The shaded purple region represents the uncertainty calculated through the propagation of error in the intensity. The dotted and solid lines show the locations for D79 and B86 respectively. The bottom left corner shows the profile of the synthesized beam (0.05'' $\times$ 0.05''; PA = 0.0\textdegree) from their parent images. \textit{(Right)} Spectral index of TW Hya's outer ring with equation \ref{Equation_spec}. The outer ring flux is calculated using the method described in Section \ref{3.4.2} and the resultant values are shown in the second last column of Table \ref{table:size_flux}.}} 
    \label{fig:spec_index}
\end{figure*}

We calculate the spectral index by plotting the flux density ($F_\nu$) of the outer ring against the frequency ($\nu$) of various ALMA bands. The spectral index ($\alpha$) is given by  
\vspace{2mm}
\begin{equation} \label{Equation_spec}
F_\nu \propto \nu^{\alpha}. \\ 
\end{equation}

{At 3.1 mm, 2.1 mm, 1.3 mm and 0.87 mm}, the flux density of the outer ring is calculated as the difference between the flux densities when using masks on their respective images at both sides of the ring location such that the total width is twice the difference between the outer ring location and the outer gap location. This is done with the \texttt{imstat} task in CASA. {At 0.65 mm, the flux density is calculated by integrating the flux in the outermost Gaussian of the model disk (see Section \ref{3.2.2} for further details).} The results of the same are shown in {the second last column of} Table \ref{table:size_flux} and Figure \ref{fig:spec_index} {(right)}. \\ 

\noindent
We also plot the 0.87 mm - 1.3 mm spectral index radial profile\footnote{We do not plot the 0.65 mm - 1.3 mm spectral index profile as the substructures are not resolved {in the image-plane} at our current resolution of the 0.65 mm observation. We detect emission in the outer region where we expect B86 to be present (see section \ref{3.2}) but we do not have enough resolution to compare in scales of 5 au.} {using the profiles in Figure \ref{fig:macias_profiles}} with the following equation - 
\vspace{2mm}
\begin{equation} \label{Equation_spec_profile}
\alpha = \frac{log\:(I_{\nu1}/I_{\nu0})}{log\:(\nu_{1}/\nu_{0})}. \\
\end{equation}
\vspace{2mm}

The uncertainties in the spectral index were calculated through the propagation of error in the intensity. This has also been shown in Figure \ref{fig:spec_index} {(left)}. In the spectral index radial profile, the sharp peak at 79 au coincides with D79. A sharp drop at $\sim$ 83 au is also consistent with B86 to the noise level. Similar to the results of previous studies, the spectral index decreases at the continuum ring and increases at the continuum gap. Our results also align with previous studies where they noticed a generally increasing trend in spectral index with an increase in radius. The spectral index profiles of TW Hya's inner 60 au \citep{tsukagoshi2016, huang2018a, macias2021} exhibit regions that are "flattened" at $\alpha \sim$ 2 covering several au, with values reaching as high as 4 near 80 au. We obtain an outer ring spectral index of 3.66 $\pm$ 0.14, which is higher than the spectral index of the inner regions. This may indicate dust grains that are optically thin and much smaller than 1 mm in size \citep{testi}. \citealp{natta2007} also found that spectral index values $\sim$ 3.5 are due to dust grains smaller than 1 mm. Interstellar medium grains in the (sub)micrometer range, which generally give off optically thin emission, also have a spectral index $\sim$ 3.5 - 4.0. Overall, the spectral index value is expected to increase with time due to the gradual loss of mm-sized pebbles from the outer disk. \citealp{dullemond2018} suggested that when substructures are optically thick, they are expected to reach saturation along their surface density peaks, resulting in the formation of rings with "flat-topped" profiles. The absence of flattened regions near B86 in the spectral index radial profile further indicates that the dust here is optically thin.

\subsubsection{Dust mass} \label{3.4.3}

The dust masses of the B86 ring at various frequencies were calculated under the assumption of optically thin emission, and that the dust continuum follows the laws of black body emission (\citealp{hillenbrand1992}) - 

\begin{equation} \label{Equation_mass}
M_{dust} = \frac{D^2 \: F_\nu}{\kappa_\nu \: B\:(\nu, T_{dust}) }, \\ 
\end{equation}

\vspace{1mm}
\noindent
where, $B\:(\nu, T_{dust})$ is given by the Planck radiation law \\
\begin{equation} \label{Equation_planck}
B\:(\nu, T_{dust}) =\frac{2 h\nu^3}{c^2}\frac{1}{ e^{\frac{h\nu}{kT_{dust}}}-1}.
\end{equation}

For the dust opacity, we assumed the same as \citealp{andrews2013}, where $\kappa_\nu$ = 2.3 cm$^2$ g$^{-1}$ $\times$ ($\nu\:$/ 230 GHz)$^{0.4}$, $T_{dust}$ = 20 K and D is the distance from the TW Hya disk in pc. The flux $F\nu$ was scaled with respect to D. The resulting ring masses are shown in the last column of Table \ref{table:size_flux}. \\

Considering the total flux of the 0.65 mm disk to be 3075.11 $\pm$ 71.55 mJy (see Section \ref{3.2.2}), we get a total dust mass of 82.55 $\pm$ 1.92 $M_\oplus$. {This is consistent with the dust mass ($\sim$ 80 $M_\oplus$) from \citealp{andrews2016_twhya} (after correcting to the new GAIA distance).} When using the total disk fluxes in Table \ref{table:1}, the 0.87 mm disk mass can be estimated to be 70.12 $\pm$ 0.04 $M_\oplus$. This estimate is close to the 0.87 mm disk mass (54 $\pm$ 5 $M_\oplus$) calculated by \citealp{macias2021} using the standard dust opacity law. Our estimate is a factor of 3-4 less than the disk mass calculated by \citealp{macias2021} when they considered the effects of self-scattering and optical depths. They suggested that the high optical depth of the disk up to R $\sim$ 55 au is likely the reason for this discrepancy. Our 0.65 mm B96 dust mass estimate (1.59 $\pm$ 0.39 $M_\oplus$) indicates that this ring contains upto $\sim$ 2\% of the total disk mass. The B86 dust masses at 0.87 mm (2.04 $\pm$ 0.02 $M_\oplus$) and at 1.3 mm (1.08 $\pm$ 0.01 $M_\oplus$) are estimated to be similar to the B96 dust mass at 0.65 mm.

\section{Discussion} \label{4}

\subsection{Why didn't we detect B86 before?} 

Extended continuum emission in disks can be challenging to detect due to low {surface} brightness levels. Therefore, in order to resolve outer disk substructures, we not only require a high angular resolution but also a high continuum sensitivity. \citealp{rosotti2019} showed that at ALMA Band 7 ($\sim$ 0.87 mm), a sensitivity of $\sim$ 0.001 K is required to detect viscous spreading in disks. They also suggested that larger disk size estimates such as the R$_{95}$ are preferable to the commonly used R$_{68}$ when representing emission in extended dust disks. Similarly, \citealp{ilee} suggests that R$_{99}$ is a better disk size estimate as R$_{90}$ may be missing important disk structure. \\ 

The 0.65 mm observation is better suited to study the outer disk region, as its shorter wavelength makes it more sensitive to the emission of smaller dust grains. At our current angular resolution ($\sim$ 0.5'', 30 au), we detect emission at R $\geq$ 80 au {with parametric modeling} and also propose the locations for possible substructures (i.e. D86 and B96) with an accuracy of $\sim$ 0.2 times our angular resolution. We also find the substructures D79 and B86 at 0.87 mm and 1.3 mm {with image-plane retrieval}. This aligns closely with the outer disk substructures (D82 and B91) proposed by \citealp{ilee}. They emphasize that a gap at $\sim$ 85 au has been consistently found in scattered light as well as molecular lines \citep{boekel2017, debes2017}. In agreement to their suggestions, we find that the R$_{99}$ disk size estimates are better suited to describe the emission in the outer regions of TW Hya, especially at shorter wavelengths where the {surface} brightness levels are high. For our 0.87 mm and 1.3 mm observations, we find that smoothing the images with circular beams increases the S/N by a factor of $\sim$ 1.5 and decreases the rms by half. The S/N is especially important in the low {surface} brightness outer regions where the noise levels are high. \\

\subsection{Origin of the outer ring emission} 



The high value of the spectral index in the outer ring ($\sim$ 3.7) implies that the dust grains are much smaller than the emitting wavelength. Our results indicate that the B96 ring at 0.65 mm may be wider than the B86 ring at 0.87 mm and 1.3 mm. A sharp drop in the spectral index profile at $\sim$ 83 au (see Figure \ref{fig:spec_index} {(left)}), where we expect B86 to be present, could indicate grain growth. \citealp{pinilla2012} found that pressure bumps can reduce the effect of radial drift and allow millimeter sized dust grains to be retained in the form of rings, which can be {observed with} ALMA. This means that the outer ring (i.e. B86) in the extended dust disk of TW Hya may be accompanied by a pressure bump to prevent the rapid drainage of dust grains in the outer disk. {Due to the low inclination of TW Hya, it is difficult to detect pressure bumps with kinematic variations since these observations rely on the deviations of the keplerian motion in the azimuthal direction of the disk (e.g., HD163296 in \citealp{izquierdo2023}).} Since we do not find the same substructure at 2.1 mm and 3.1 mm, it is unclear whether the said pressure bump would also trap larger sized mm dust grains. According to the models simulated by \citealp{long2020}, if pressure bumps are responsible for effectively trapping dust particles in outer disks, the disk sizes should not vary at different wavelengths. If the larger dust grains are indeed absent in the R $\geq$ 60 au regions of TW Hya, a possible explanation would be that these dust grains migrated inwards first and then the pressure bumps were formed \citep{pinilla2015}. \citealp{macias2021} also suggested that the rings in TW Hya within 20 au $\leq$ R $\leq$ 60 au could be accompanied by annular gas pressure bumps causing millimeter and centimeter sized grains to be accumulated. These rings would then be suitable locations for streaming instability to trigger planetesimal formation. \\

Although we do not have enough resolution to compare in scales of 10 au, a possible explanation for the B96 peak at 0.65 mm to be shifted further away from the B86 peak at 0.87 mm and 1.3 mm could be a dust "traffic jam". If we assume that B86 is the site for a pressure maxima and the small dust grains are well coupled to the gas, we could expect a traffic jam to slow down the inward drift of particles \citep{pinilla2016}. The small dust grains could thus, be spread out and have its peak shifted away from the pressure maxima.

\section{Conclusions} \label{5}

We present {archival} $\sim$ 0.5 arcsecond angular resolution ALMA observations of the continuum emission of the protoplanetary disk around TW Hya at a wavelength of 0.65 mm. We analyze this data together with high angular resolution observations that were processed by \citealp{macias2021} at the wavelengths of 0.87 mm, 1.3 mm, 2.1 mm and 3.1 mm. {With image-plane retrieval}, our 0.87 mm and 1.3 mm observations resolve substructures in the R $\geq$ 60 au of TW Hya, namely D79 and B86. At a wavelength of 0.65 mm, we also find a quasi-constant emission between 80 and 100 au by fitting a parametric model to the visibilities. Our results suggest the presence of the substructures D86 and B96 with {constraints that are $\sim$ 3 times better than} the spatial resolution ($\sim$ 30 au). These substructures are coincident with the D82 and B91 proposed by \citealp{ilee}, who also pointed out that the outer disk gap has been identified before in scattered light and molecular lines. Due to lower sensitivity, we do not confirm the existence of this outer disk emission at 2.1 mm and 3.1 mm. To resolve the 0.65 mm outer disk substructures {in the image-plane}, we would require high angular resolution observations of at least $\sim$ 0.1 arcsecond (6 au) with {high} sensitivity and S/N. \\

The multiwavelength observations have allowed us to constrain the properties of the outer disk emission. We believe that the R$_{99}$ disk size estimate ($\sim$ 96 au at 0.65 mm, 0.87 mm and 1.3 mm) represents the outer disk emission more accurately than the R$_{68}$ estimate. The {outer} ring at 0.65 mm may be wider than the {outer} ring at 0.87 mm and 1.3 mm. We find that the B86 ring has a spectral index of $\sim$ 3.7, indicating optically thin small dust grains. Optically thin emission is further supported by the absence of flattened regions in the 0.87 mm - 1.3 mm spectral index radial profile. We also obtain a sharp drop at $\sim$ 83 au in the spectral index radial profile {which is consistent with the B86 ring peak to the noise level}. This would indicate grain growth and support the notion of B86 being a "dust trap," which may be accompanied by a pressure bump. Another possible origin for the outer ring may be a "traffic jam" where the inward drift of dust grains has slowed down. {With parametric modeling}, we estimate the outer ring to have a flux density of $\sim$ 60 mJy at 0.65 mm. We also estimate a mass of $\sim$ 1.59 $M_\oplus$ and therefore, it may account for up to 2\% of the total dust mass at 0.65 mm. \\

\bibliographystyle{aa} 
\bibliography{ref}

\begin{acknowledgements}
We are grateful to Enrique Macias for making his data available. We thank the referee, John Ilee, for an insightful report. SD acknowledges support from the DAAD-WISE scholarship (grant number 57655004). This paper makes use of the following ALMA data: ADS/JAO.ALMA\#2016.1.01399.S, JAO.ALMA\#2015.1.00686.S, JAO.ALMA\#2015.1.00845.S, JAO.ALMA\#2015.A.00005.S, JAO.ALMA\#2016.1.00173.S, JAO.ALMA\#2016.1.00229.S, JAO.ALMA\#2016.1.00629.S, JAO.ALMA\#2016.1.00842.S, JAO.ALMA\#2016.1.01158.S, JAO.ALMA\#2017.1.00520.S, and JAO.ALMA\#2018.1.01218.S. ALMA is a partnership of ESO (representing its member states), NSF (USA) and NINS (Japan), together with NRC (Canada), MOST and ASIAA (Taiwan), and KASI (Republic of Korea), in cooperation with the Republic of Chile. The Joint ALMA Observatory is operated by ESO, AUI/NRAO and NAOJ. The National Radio Astronomy Observatory is a facility of the National Science Foundation operated under cooperative agreement by Associated Universities, Inc.
\end{acknowledgements}

\begin{appendix} 

\section{Intensity profiles generated with frank} \label{appendix_A}

\begin{figure*}
    \centering
    \vspace{5cm}
    \includegraphics[width=0.49\textwidth]{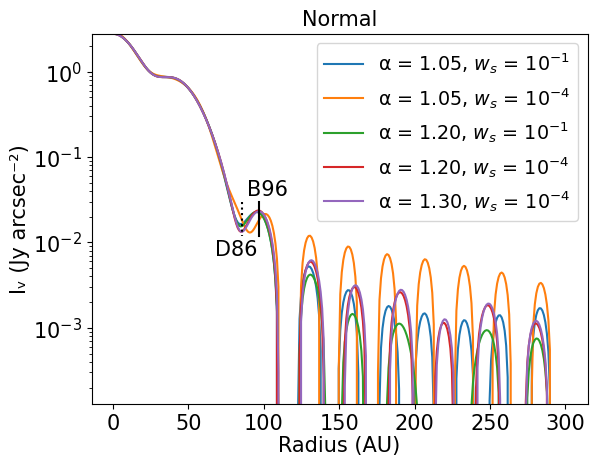}
    \includegraphics[width=0.49\textwidth]{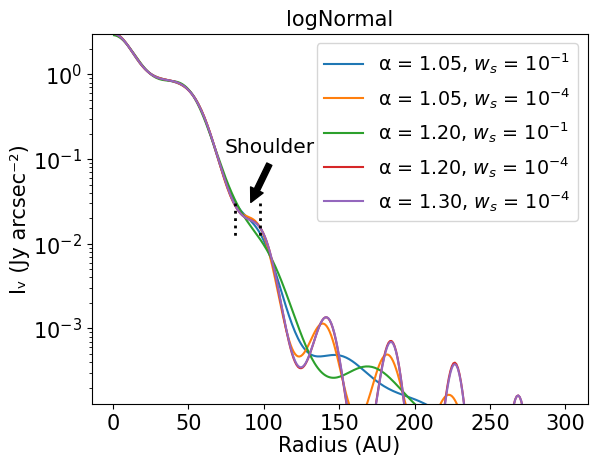} 
    \vspace{1.5mm}
    \caption{\textit{(Left)} Brightness profiles reconstructed with frank's "Normal" method using different $\alpha$ and w$_{smooth}$ values. We get a dip and a peak resembling substructures, which are labelled as D86 and B96 respectively. These are shown with dotted and solid lines. \textit{(Right)} Brightness profiles reconstructed with frank's "logNormal" method using the same $\alpha$ and w$_{smooth}$ values as the plot on the left. We get a shoulder-like emission between 80 au and 97 au. The extent of the shoulder is shown with dotted lines. The intensity axis is cut off at the rms of the 0.65 mm observation (0.14 mJy arcsec$^{-2}$).} 
    \label{fig:frank_normal}
\end{figure*}

We construct brightness profiles for the continuum emission at 0.65 mm with the \texttt{Frankenstein} code \citep{tazzari2017}. We vary the $\alpha$ and w$_{smooth}$ values as $\alpha$ = \{1.05, 1.20, 1.3\} and w$_{smooth}$ = \{10$^{-1}$, 10$^{-4}$\}. The "Normal" method fits all brightness values with non-parametric Gaussians, whereas the "logNormal" method avoids fitting the negative brightness values. This is shown in Figure \ref{fig:frank_normal}. An advantage of the logNormal method is that it produces fewer noise-induced artifacts. With the Normal method, we get a dip and a peak at 86 au and 96 au respectively which resemble substructures. Particularly, the $\alpha$ parameter causes a shift in the ring peak by $\sim$ 5\%. With the logNormal method, we no longer get a sharp dip and a peak, but the outer ring appears in the form of a shoulder between 80 au and 97 au. A higher w$_{smooth}$ causes significant smoothing of this shoulder. 

\section{MCMC fitting and results} \label{appendix_B}

\vspace{1.5mm}
As discussed in Section \ref{3.2.2}, we model the 0.65 mm disk around TW Hya with MCMC (emcee package; \citealp{emcee}) and use \texttt{galario} \citep{tazzari2018} to generate synthetic visibilities of the 2D model image. Using equation \ref{Equation1}, we get 3 Gaussians - 1 central emission (I$_0$, $R_0$ = 0, $\sigma_0$) and 2 rings (I$_1$, $R_1$, $\sigma_1$ and I$_2$, $R_2$, $\sigma_2$) . We also fit the right ascension offset ({$\Delta$R}) and declination offset ({$\Delta$D}). By default, galario shifts {$\Delta$R} > 0 towards East and {$\Delta$D} > 0 towards North. This gives us a total of 10 parameters. We run our MCMC sampler with 80 walkers, for a total of 10,000 steps after a burn-in of 2000 steps. We assume our parameters to have the following uniform priors : \\
log (I$_0$ Jy/steradian) = [-7, 2]; $\sigma_0$ arcsec = [0.03'', 0.83'']; log (I$_1$ Jy/steradian) = [-7, 2]; $R_1$ arcsec = [0.03'', 1.16'']; $\sigma_1$ arcsec = [0.03'', 0.83'']; log (I$_2$ Jy/steradian) = [-7, 2]; $R_2$ arcsec = [1.16'', 1.66'']; $\sigma_2$ arcsec = [0.03'', 0.83'']; {$\Delta$R} arcsec = [-2'', 2'']; {$\Delta$D} arcsec = [-2'', 2''] \\

\begin{figure*}
    
    \centering
    \includegraphics[width=1.03\textwidth]{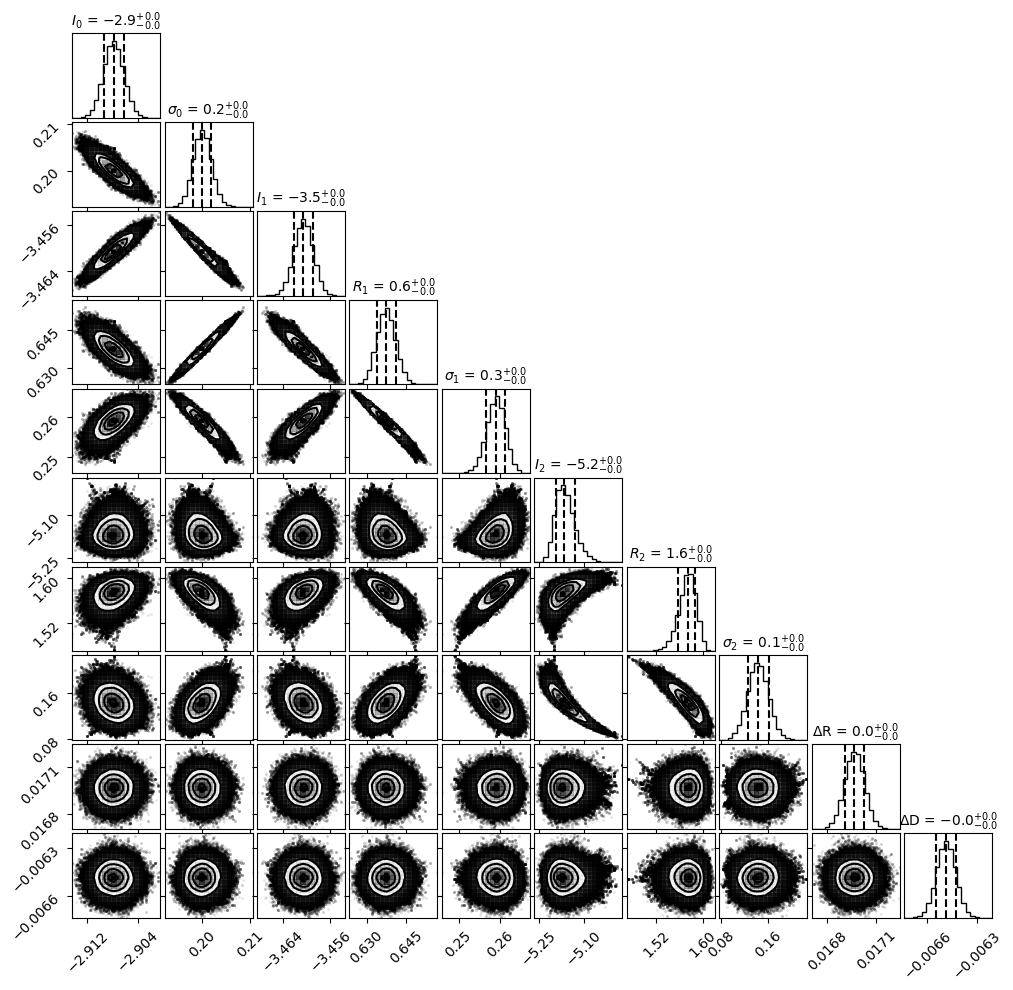}
    \caption{{The staircase plot of the MCMC samples (see Section \ref{3.2.2}) where the top row shows the 50th percentile values and the uncertainties are calculated from the 16th and 84th percentiles respectively.}} 
    \label{fig:A.2}
\end{figure*}

The results of the best-fit model (the model with the maximum likelihood) are shown in Table \ref{table:A.1}. The staircase plot of the chains of the best-fit model is shown in Figure \ref{fig:A.2}. The top row of each column contains the median (50th percentile) of the fitted parameters correct to two decimal places. The uncertainties are calculated from the 16th and 84th percentiles of the marginalized distributions.

\begin{table*}
\vspace{4mm}
\caption{Results of the MCMC fitting from the best-fit model of TW Hya at 0.65 mm}         
\label{table:A.1}      
\centering   
\begin{tabular}{c c c c c c c c c c}     
\hline\hline       
\thead{log (I$_0$) \\ (Jy/steradian)} &  \thead{$\sigma_0$ \\ (au)} & \thead{log (I$_1$) \\ (Jy/steradian)}  &  \thead{ $R_1$ \\ (au)} &  \thead{$\sigma_1$ \\ (au)} &  \thead{log (I$_2$) \\ (Jy/steradian)}  &  \thead{ $R_2$ \\ (au)} &  \thead{$\sigma_2$ \\ (au)} & \thead{{$\Delta$R} \\ (au)} & \thead{{$\Delta$D} \\ (au)} \\ [3ex]

\midrule
   -2.91$^{+0.00}_{-0.00}$ & 11.99$^{+0.11}_{-0.11}$ & -3.46$^{+0.00}_{-0.00}$ & 38.26$^{+0.22}_{-0.21}$ & 15.62$^{+0.13}_{-0.14}$ & -5.17$^{+0.04}_{-0.03}$ & 94.86$^{+0.81}_{-0.97}$ & 8.38$^{+1.07}_{-1.04}$ & 1.02$^{+0.01}_{-0.01}$ & -0.39$^{+0.01}_{-0.01}$ \\
   
\midrule     
\end{tabular}
\vspace{4mm}

\begin{tabular}{c c c c c }     
\hline\hline       
\thead{R$_{68}$ (au)} &  \thead{R$_{95}$ (au)} & \thead{R$_{99}$ (au)}  &  \thead{ B96 Flux \\ Density (mJy)} &  \thead{Total Flux \\ Density (mJy)}\\ [3ex]

\midrule
\vspace{2mm}
   46.85$^{+0.61}_{-0.61}$ & 69.13$^{+1.20}_{-1.20}$ & 95.00$^{+3.00}_{-3.00}$ & 59.24$^{+14.64}_{-10.31}$ & 3075.11$^{+71.55}_{-68.44}$ \\
   
\midrule     
\end{tabular}

\end{table*}

\section{Procedure for generating azimuthally averaged intensity profiles from images} \label{appendix_C}

The one-dimensional intensity profiles from two-dimensional images are generated with the following procedure - \\

\begin{enumerate}[label=(\alph*)]
\item A model ellipse is created based on TW Hya's geometry which maps out the disk's radial points. This is done with a rectangular grid created with the image's pixel-size and image-size. Any offsets between the center of the image and that of the model are corrected manually. The grid is then inclined and rotated to mirror its projection in the sky-plane. For this, the inclination angle and position angle are set to 5.8\textdegree \space and 151.6\textdegree \space respectively \citep{huang2018a}. Finally, the cartesian coordinates are transformed to polar coordinates. \\
\item Next, we consider a radial bin equal to the pixel-size of the model and azimuthally average the brightness of all the pixels lying within this bin. We also calculate an equivalent radius by averaging all the radial points within the bin. This process is repeated until we reach the end of the model ellipse. \\
\item The uncertainty (u) in the emission is also calculated corresponding to each azimuthally averaged brightness value. This can be given by the rms of the image (rms) and the number of elements in a ring (n). The latter is dependent on the major and minor axes of the beam ($maj_{b}$ and $min_{b}$) with which the image was convolved, the radial bin (dr) and the inner radius of each ring (r).  \\

\begin{equation} \label{Equation2}
n = \frac{A_{ring}}{A_{beam}} = \frac{\pi\:((r + dr)^2 - r^2)}{\pi\:\times\: maj_{b}\:\times\: min_{b}}, \\ \\
\end{equation} \\
and
\begin{equation} \label{Equation3}
u = \frac{rms}{2\sqrt{n}}. \\
\end{equation}

\end{enumerate}

\end{appendix}

\end{document}